# Hybrid Precoding for Multi-Group Physical Layer Multicasting

Meysam Sadeghi*, Luca Sanguinetti†, and Chau Yuen*
* Singapore University of Technology and Design, Singapore
† University of Pisa, Dipartimento di Ingegneria dellInformazione, Italy
Email: meysam@mymail.sutd.edu.sg, luca.sanguinetti@unipi.it, and yuenchau@sutd.edu.sg

*Abstract*—Next generation of wireless networks will likely rely on large-scale antenna systems, either in the form of massive multi-input-multi-output (MIMO) or millimeter wave (mmWave) systems. Therefore, the conventional fully-digital precoders are not suitable for physical layer multicasting as they require a dedicated radio frequency chain per antenna element. In this paper, we show that in a multi-group multicasting system with an arbitrary number of transmit antennas, $G$ multicasting groups, and an arbitrary number of users in each group, one can achieve the performance of any fully-digital precoder with just $G$ radio frequency chains using the proposed hybrid multi-group multicasting structure.

*Index Terms*—Hybrid precoding, multi-group multicast precoding, physical layer multicasting, massive MIMO, mmWave.

## I. INTRODUCTION

There has been a tremendous growth in the data traffic of wireless networks, and it is anticipated to continue in the upcoming years [1]. A considerable part of this traffic is of interest to groups of users rather than a single user, e.g., live broadcasting of sporting events, regular updates, and news headlines. To address such a traffic demand, physical layer multicasting was introduced in [2]. In particular, precoding design for physical layer multicasting has been extensively studied in the literature, e.g. [2]–[8] and references therein.

The main idea in physical layer multicasting is to employ the channel state information at the transmitter to design precoding vectors that optimizes a desired utility function given some constraints. In this context two classes of problems are of interest, the so-called quality of service (QoS) and max-min fairness (MMF) problems. In QoS problem, the precoding vectors are designed to minimize the power consumption at the transmitter while meeting predefined quality of service constraints. In MMF problem, the precoding vectors are designed to maximize the minimum of a desired objective while meeting a power consumption constraint.

The precoding design for the QoS and MMF problems in single group multicasting are studied in [2]. Both problems are then extended to multi-group multicasting in [3]. The single group MMF problem for massive multi-input-multi-output (MIMO) is studied in [4], and its extension to multi-group systems is investigated in [5]. The MMF problem is then revisited under per antenna power constraint in [6]. Recently low computational complexity algorithms for massive MIMO multicasting is presented in [7]. These new algorithms not only reduce the computational complexity but also significantly outperforms the methods in [2]–[6].

All the aforementioned works are based on fully-digital (FD) precoding schemes. Although an FD precoder enables us to design the precoding vectors ideally, it is not desirable for future wireless networks. These networks employ a very large number of antennas at the base station (BS) in the form of massive MIMO or millimeter wave (mmWave) systems [9]–[14], and as an FD precoder would require a dedicated baseband and RF chain for each antenna, FD precoders become less practical due to their high cost and power consumption [15].

To address this practical constraint, hybrid precoding was presented in [16]. A hybrid precoder is made up of a low-dimensional baseband precoder followed by a high-dimensional radio frequency (RF) precoder. The RF precoder is fully implemented by cost efficient analog phase shifters, which in turn enforces an extra modulus constraint on the design of the precoding vectors. Using the basis pursuit principle, an algorithm for hybrid precoding in point-to-point MIMO systems is presented in [15]. Hybrid precoding for both point-to-point and multi-user MIMO is studied in [16]. A low-complexity hybrid precoding for massive MIMO system is presented in [17].

All of the aforementioned works deal with hybrid precoding for unicast transmissions. On the other hand, hybrid precoding for physical layer multicasting has received little attention so far. The first attempt can be found in [18], where the authors focus on single-group multicasting for mmWave communications and present a hybrid precoding scheme achieving the same performance of the FD precoder with only $\sum_{i=1}^{K} L_i$ RF chains, where $K$ denote the number of user equipments (UEs) within the group and $L_i$ accounts for the number of multipath components of the channel of UE $i$. Therein, a heuristic solution requiring a smaller number of RF chains is also proposed for the MMF problem.

The aim of this paper is to extend the work in [18] and to reveal how many RF chains are needed in a multi-group multicasting system to achieve the same performance of an FD precoder. In particular, we show that if there exist $G$ multicasting groups, then $G$ RF chains are required to perfectly implement any FD precoder. This holds true independently of the number of UEs in each group and of the number of antennas employed at the BS. Interestingly, this implies that in single-group multicasting, just a single RF chain is enough to perfectly implement any FD precoder, which is in sharp contrast with [18]. Another interesting aspect of the proposed solution is that it is problem independent, while in the literature the hybrid precoder is usually designed for a specific objective function (e.g. sum rate maximization, or fairness) and cannot be applied elsewhere.

The rest of the paper is organized as follows. Section II introduces the system model and the problem under investigation. The proposed hybrid precoding scheme is illustrated in Section III. Section IV shows some numerical results whereas conclusions are drawn in Section V.

*Notations*: The following notation is used throughout the paper. Scalars are denoted by lower case letters whereas boldface lower (upper) case letters are used for vectors (matrices). The transpose, conjugate transpose, and rank of a matrix are denoted by $(.)^T$, $(.)^H$, and $\mathrm{rank}(.)$. A circular symmetric complex Gaussian random variable $x$ with zero mean and variance $\sigma^2$ is denoted by $x \sim \mathcal{CN}(0, \sigma^2)$.

## II. SYSTEM MODEL AND PROBLEM DESCRIPTION

Consider a single-cell large-scale antenna array system in which a BS, equipped with $N_{\mathrm{RF}}$ RF chains and $N$ antennas with $N > N_{\mathrm{RF}}$, serves $G$ multicasting groups. Denote by $\{1, \ldots, G\}$ the set of indices of all groups and $\mathcal{K}_j$ the set of indices of UEs associated with group $j$, with cardinality $K_j = |\mathcal{K}_j|$ and such that $\mathcal{K}_j \cap \mathcal{K}_i = \emptyset$, $j \neq i$, i.e., each UE is associated with a single group. Within this setting, the goal of physical layer multicasting is to design (according to some optimal criterion) the FD precoding matrix $\mathbf{W}_{\mathrm{FD}} = [\mathbf{w}_1, \ldots, \mathbf{w}_G] \in \mathbb{C}^{N \times G}$ with $\mathbf{w}_j \in \mathbb{C}^N$ being the precoding vector of group $j$. However, this would require $N$ RF chains, which is not desirable when $N$ takes large values, as envisioned in future wireless networks [9], [11]. To overcome this issue, hybrid precoding aims at designing an analog-digital precoder $\mathbf{W}_{\mathrm{HP}} \in \mathbb{C}^{N \times G}$ such that

$$\mathbf{W}_{\mathrm{HP}} = \mathbf{W}_{\mathrm{RF}} \mathbf{W}_{\mathrm{BB}} \quad (1)$$

where $\mathbf{W}_{\mathrm{BB}} \in \mathbb{C}^{N_{\mathrm{RF}} \times G}$ is the baseband precoder and $\mathbf{W}_{\mathrm{RF}} \in \mathbb{C}^{N \times N_{\mathrm{RF}}}$ is the analog one. Due to the practical constraints, $\mathbf{W}_{\mathrm{RF}}$ should be implemented solely by phase shifters. The phase shifters can have an arbitrary phase but a constant modulus, e.g. a phase shifter can be modeled as $e^{j\phi}$ with $\phi \in [0, 2\pi]$ [15], [16], [19].

Two approaches basically exist for the design of hybrid precoders. The first one aims at directly designing the hybrid precoder for the specific problem at hand (e.g. [15], [16]). The second one is problem-independent and looks for the minimum number of RF chains that are required to implement any FD precoder $\mathbf{W}_{\mathrm{FD}}$ with a hybrid one $\mathbf{W}_{\mathrm{HP}}$ [20]. The second approach is followed next.

## III. HYBRID PRECODING FOR MULTI-GROUP MULTICASTING

Let us start assuming that $\mathbf{W}_{\mathrm{FD}}$ is full rank, i.e. $r = \mathrm{rank}(\mathbf{W}_{\mathrm{FD}}) = G$. From (1), it readily follows that $\mathrm{rank}(\mathbf{W}_{\mathrm{HP}}) \leq N_{\mathrm{RF}}$, therefore at least $G$ RF chains are needed to perfectly implement $\mathbf{W}_{\mathrm{FD}}$ with $\mathbf{W}_{\mathrm{HP}}$. If $\mathbf{W}_{\mathrm{FD}}$ is a rank deficient matrix, i.e. $r = \mathrm{rank}(\mathbf{W}_{\mathrm{FD}}) < G$, then following [19] we can rewrite it as $\mathbf{W}_{\mathrm{FD}} = \mathbf{A}^{N \times r} \mathbf{B}^{r \times G}$ and find a hybrid precoder for the full rank matrix $\mathbf{A}^{N \times r}$ as $\mathbf{A}^{N \times r} = \mathbf{W}_{\mathrm{RF}}^{N \times N_{\mathrm{RF}}} \mathbf{W}_{\mathrm{BB}}^{N_{\mathrm{RF}} \times r}$ with $r$ RF chains. In this case, the digital and analog parts of the hybrid precoder of $\mathbf{W}_{\mathrm{FD}}$ are respectively given by $\mathbf{W}_{\mathrm{BB}} \mathbf{B}$ and $\mathbf{W}_{\mathrm{RF}}$. Therefore, the minimum number of RF chains required for perfectly implementing any FD precoding matrix is equal to $\min(r, G)$.

Inspired by [16] and [19], in the sequel we propose a multicast hybrid precoding structure that proves having $\min(r, G)$ RF chains is not only necessary but also sufficient to fully implement any FD multi-group multicasting precoder. To this end, we assume that $\mathbf{W}_{\mathrm{FD}}$ is given and denote its $(n, j)$th entry as $[\mathbf{W}_{\mathrm{FD}}]_{nj} = \alpha_{nj} e^{j\theta_{nj}}$ with $\alpha_{nj} \geq 0$. Then, we let $\mathbf{W}_{\mathrm{RF}} = [\mathbf{w}_{\mathrm{RF},1}, \ldots, \mathbf{w}_{\mathrm{RF},G}] \in \mathbb{C}^{N \times G}$ with

$$\mathbf{w}_{\mathrm{RF},j} = \mathbf{a}_j + \mathbf{b}_j \quad (2)$$

where $[\mathbf{a}_j]_n = e^{j\varphi_{nj}}$ and $[\mathbf{b}_j]_n = e^{j\phi_{nj}}$. We also assume that $\mathbf{W}_{\mathrm{BB}} \in \mathbb{C}^{G \times G}$ is diagonal and given by

$$\mathbf{W}_{\mathrm{BB}} = \mathrm{diag}(\rho_1, \ldots, \rho_G) \quad (3)$$

with $\rho_j > 0$. Then, the aforementioned hybrid precoder can fully implement $\mathbf{W}_{\mathrm{FD}}$ if

$$\alpha_{nj} e^{j\theta_{nj}} = \rho_j (e^{j\varphi_{nj}} + e^{j\phi_{nj}}) \quad \forall n, j. \quad (4)$$

Since $G \leq N$, the above system of equations is underdetermined (more unknowns than equations). This means that there exists an infinite number of values of $(\rho_j, \varphi_{nj}, \phi_{nj})$ satisfying (4). However, we are only interested in finding a possible solution among the infinite number of solutions. To this end, we first observe that (4) is feasible only if $\rho_j \geq \frac{1}{2} \max_n \alpha_{nj}$. This is simply because the maximum value of the left hand side of (4) is equal to $\max_n \alpha_{nj}$ and the maximum of the right hand side of (4) is equal to $2\rho_j$. In order to reduce the power used by the digital precoder we set $\rho_j$ as

$$\rho_j = \frac{1}{2} \max_{n \in \mathcal{K}_j} \alpha_{nj} \quad \forall j \in \mathcal{G}. \quad (5)$$

One could also set $\rho = \rho_j \, \forall j$, where $\rho = \frac{1}{2} \max_j \max_n \alpha_{nj}$ which increases the power consumption but simplifies the digital precoder to just one multiplier. Considering (5) as the metric then, (4) is satisfied if[1]

$$\varphi_{nj} = \theta_{nj} + \cos^{-1}\left(\frac{\alpha_{nj}}{2\rho_j}\right) \quad \forall n, j \quad (6)$$

$$\phi_{nj} = \theta_{nj} - \cos^{-1}\left(\frac{\alpha_{nj}}{2\rho_j}\right) \quad \forall n, j. \quad (7)$$

---
[1]To prove (6) and (7) (or (8) and (9)), decompose both sides of (4) to their respective real and imaginary parts and then solve the two resulting equations.

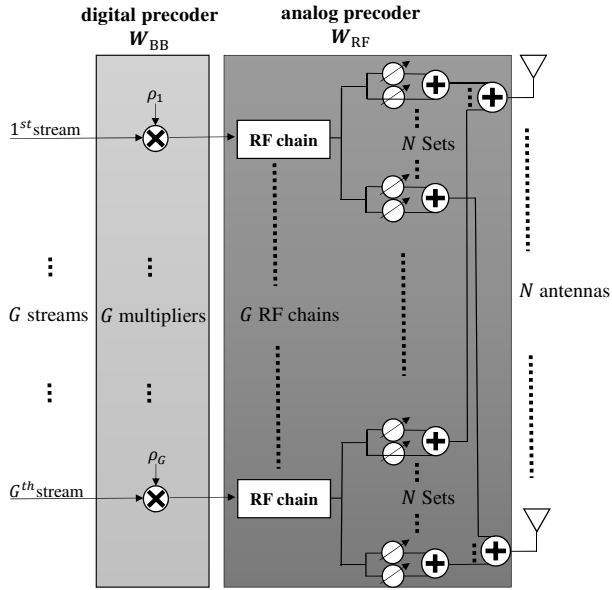

Fig. 1: Structure of the proposed hybrid precoder for multi-group physical layer multicasting.

accounting for the number of multipath components of the channel of UE $i$ and $K$ is the number of all UEs in the system. Note that $\sum_{i=1}^{K} L_i$ is substantially larger than 1. In the simplest case in which $L_i = 1 \ \forall i$ (as for example in mmWave systems), we have that $N_{\text{RF}} = K$, which is still much larger than 1.

As another special case of our set-up, if $K_j = 1 \ \forall j$ is considered, our system reduces to a downlink multi-user multi-input-single-output system. Therefore, Proposition 1 shows that for such systems any FD precoder is fully implementable with as many RF chains as the number of UEs in the system. This coincides with the results achieved in [20]. Compared to [20], our proposed structure enforces a very simple digital precoder part ($\mathbf{W}_{BB}$) that can be implement by just $G$ multipliers (or even just 1 multiplier where $\rho = \frac{1}{2} \max_j \max_n \ \alpha_{nj}$), due to (3) and (5).

Considering (5), another possible solution for (4) is as follows:

$$\varphi_{nj} = \theta_{nj} - \cos^{-1}\left(\frac{\alpha_{nj}}{2\rho_j}\right) \quad \forall n, j \quad (8)$$

$$\phi_{nj} = \theta_{nj} + \cos^{-1}\left(\frac{\alpha_{nj}}{2\rho_j}\right) \quad \forall n, j. \quad (9)$$

*Remark*: As each of the above sets of solutions, either (6) and (7) or (8) and (9), satisfy all the $N \times G$ equations in (4), there is no need for decomposing the analog precoder into more than two terms in (2) as it just increases the number of required phase shifters while do not improve the performance.

In summary, the implementation of the proposed hybrid precoder requires first to solve the problem at hand to obtain the desired $\mathbf{W}_{\text{FD}}$. Then, its entries $\{[\mathbf{W}_{\text{FD}}]_{n,j} = \alpha_{nj} e^{j\theta_{nj}}\}$ may be used to find the optimal $\mathbf{W}_{\text{RF}}$ and $\mathbf{W}_{\text{BB}}$ according to (6)-(7) or (8)-(9). Fig. 1 illustrates the block diagram of the proposed hybrid multi-group multicasting precoder. As seen, it requires a simple digital precoder made by $G$ multipliers, and an analog precoder with $G$ RF chains and $2GN$ phase shifters. Therefore, we have the following result.

**Proposition 1.** *Consider a single-cell multi-group multicasting system with $G$ groups, an arbitrary number of UEs per group, and an arbitrary number BS antennas. Then, the necessary and sufficient number of RF chains $N_{\text{RF}}$ required to perfectly implement any FD precoder with a hybrid precoding scheme as that illustrated in in Fig. 1 is equal to $G$.*

In the special case of single-group multicasting, Proposition 1 implies that just a single RF chain is enough to perfectly implement any FD precoder. This is a substantial saving (in terms of practical implementation) with respect to the solution presented in [18] for which $N_{\text{RF}} = \sum_{i=1}^{K} L_i$, where $L_i$

## IV. SIMULATIONS RESULTS

Numerical results are now used to assess the performance of the proposed hybrid precoder. Comparisons are made with the scheme presented in [18] and also with the corresponding FD multicasting scheme, which determines the benchmark. To emphasize that the proposed hybrid precoder is problem-independent, we consider both classes of problems that are generally investigated in physical layer multicasting, namely, the QoS and MMF problems [2], [3]. We assume a typical value of 10 Watt for the total transmit power for the MMF problem, and a requested SINR equal to 128 (in accordance with 5G requirements [12]) for the QoS problem. But note that the proposed hybrid precoder structure is valid for any values of the total transmit power and the requested SINRs. To solve the QoS or MMF problems and obtain $\mathbf{W}_{\text{FD}}$, we use the algorithm presented in [3], which employs the semidefinite relaxation technique followed by a randomization and a multicast multigroup power control policy.[2]

For the communication environment, we consider a mmWave communication system similar to [15], [17]–[19], although our results are general and independent of the communication model. Denoted by $\mathbf{h}_{jk}$ the channel between the BS and UE $k$ within group $j$. The following geometric model is adopted to capture the poor scattering behavior of mmWave communication systems [15], [17]–[19]

$$\mathbf{h}_{jk}^H = \sqrt{N/L} \sum_{l=1}^{L} \alpha_{jkl} \ \mathbf{a}^H(\phi_{jkl}, \theta_{jkl}) \quad (10)$$

where $L$ is the number of paths (for simplicity it is the same for all UEs in the system), $\alpha_{jkl} \sim \mathcal{CN}(0,1)$ is the complex gain of $l$th path of UE $k$ in group $j$, $\mathbf{a}(\phi_{jkl}, \theta_{jkl})$ is the array response vector at the azimuth and elevation of $\phi_{jkl}$ and

---

[2]For the randomization phase, 100 samples are generated using the Gaussian randomization method [3].

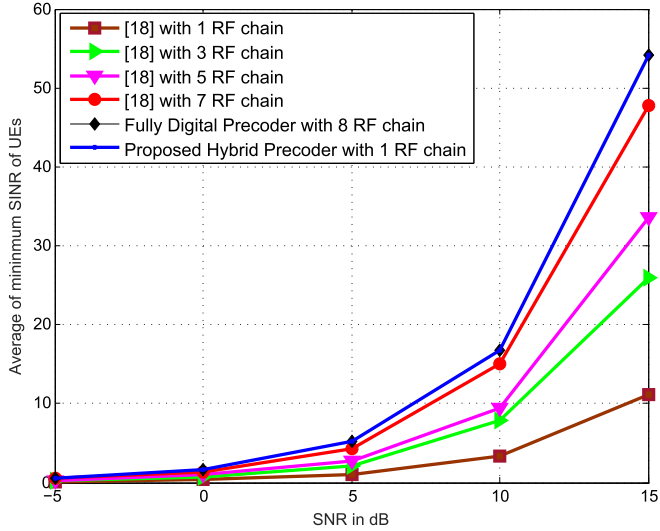

Fig. 2: MMF problem for single-group physical layer multicasting.

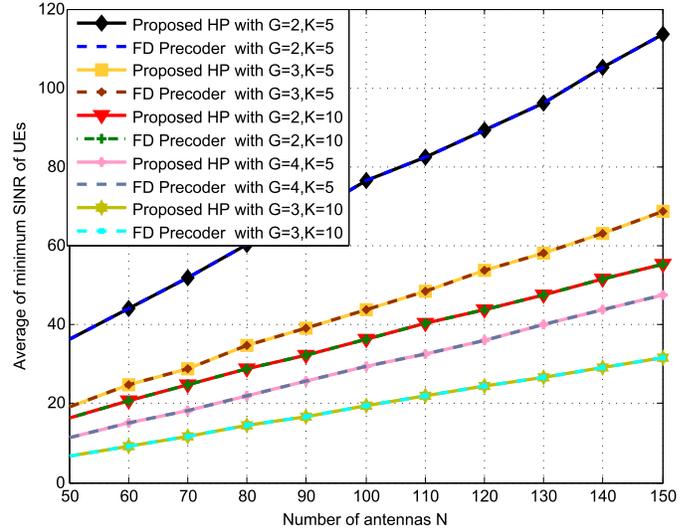

Fig. 4: Average of minimum SINR (MMF problem) versus number of antennas.

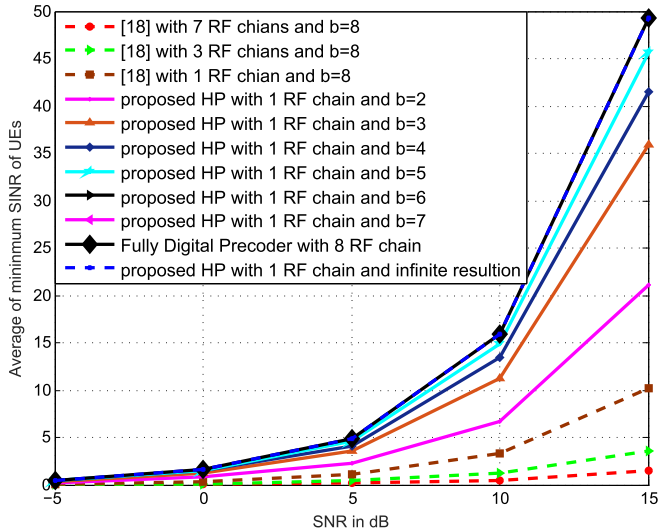

Fig. 3: MMF problem for single-group physical layer multicasting: the effect of finite resolution phase shifters.

$\theta_{jkl}$, respectively. We assume that the BS is equipped with a uniform linear array such that [18]

$$\mathbf{a}(\phi_{jkl}) = \frac{1}{\sqrt{N}}[1, e^{j\frac{2\pi}{\lambda}\Delta \sin(\phi_{jkl})}, \ldots, e^{j\frac{2\pi}{\lambda}\Delta(N-1)\sin(\phi_{jkl})}]^T$$

where $\lambda$ is the wavelength and $\Delta$ is the antenna spacing. We assume that $\Delta = \lambda/2$, $L = 3$, and the angle values $\{\phi_{jkl}; \forall j,k,l\}$ are drawn independently from a uniform distribution over $[0, 2\pi]$. This results in random locations for UEs irrespective of the group they belong.

In Fig. 2, we consider the MMF problem in a single-group multicasting system with $N = 8$ and $K = 7$ and illustrates the average of the minimum SINR of all UEs. In all simulations, the averaging is performed over 100 channel realizations. The FD precoder with $N_{\rm RF} = 8$ determines the performance upper bound. The hybrid multicast precoding in [18] is reported for different values of $N_{\rm RF}$. Interestingly, the hybrid precoder of [18] does not achieve the upper bound even with 7 RF chains while the proposed one provides the same performance as the FD precoder with $N_{\rm RF} = 1$.

The results presented in Fig. 2 for both [18] and the proposed structure are based on phase shifters with infinite resolutions. The effect of finite resolution phase shifters on the proposed hybrid multicast precoder is evaluated in Fig. 3, under the same system setup as Fig. 2. As observed, the proposed scheme can achieve more than 80% (90%) of the performance of the infinite resolution precoder with just $b = 3$ ($b = 4$) bits (for the phase shifter resolution) while the performance of [18] severely degrades even for $b = 8$. Therefore, our scheme exhibits robustness and experiences only a small degradation of performance.

Note that in Figs. 2 and 3 we considered a system with $G = 1$ and $N = 8$ antennas while in general we might have a massive MIMO multi-group multicasting system, which employs hundreds of antennas [9]. This restriction was due to the high complexity of the proposed algorithm in [18] with respect to $N$ and its single-group nature. More precisely, [18] requires to solve a combinatorial number of instances of the MMF problem, which increases very fast with the number of BS antennas $N$. On the contrary, the proposed method just need to solve the MMF problem once. This makes it particularly appealing for practical implementation. Moreover, it can handle multi-group scenarios. To verify these two aspects, Figs. 4 and 5 compare the performance of the proposed scheme with that of an FD precoder for MMF and QoS problems, considering various system setups. Note that we can achieve the same performance of the FD precoder with a much smaller number of RF chains. Consider for example a system with $G = 2$, $K = 10$, and $N = 150$. Then, the FD precoder requires $N_{\text{RF}} = 150$ RF chains while the proposed scheme achieves the same performance with just $N_{\text{RF}} = 2$ RF chains.

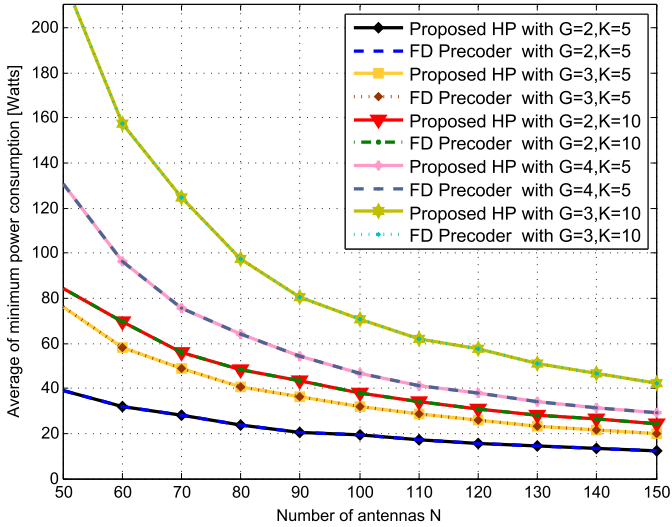

Fig. 5: Average of minimum power consumption (QoS problem) versus number of antennas.

## V. Conclusions

This paper proposed a hybrid precoding structure for multi-group physical layer multicasting that can perfectly implement any FD precoder with just as many RF chains as the number of multicasting groups in the system. This is achieved independently of the number of UEs and the number of BS antennas. In the special case of single-group multicasting, the proposed solution can perfectly implement any FD precoder with just one single RF chain, which significantly improves the existing result on hybrid precoding for physical layer multicasting. This makes it appealing for practical implementation. Moreover, the proposed approach was independent of the problem of interest as it was verified by applying it to the QoS and MMF problems.